\documentclass[11pt,a4paper]{article}

 \usepackage[english]{babel}
 \usepackage[T2A]{fontenc}
 \usepackage[cp1251]{inputenc}
 \usepackage{amsmath}
\usepackage{graphicx}
\usepackage{amssymb}
\usepackage{color}
\usepackage{amsfonts}
\usepackage{wrapfig}
\usepackage{caption}

\newcommand{\be}{\begin{equation}}
\newcommand{\ee}{\end{equation}}
\newcommand{\bea}{\begin{align}}
\newcommand{\eea}{\end{align}}
\newcommand{\bt}{\begin{tabular}}
	\newcommand{\et}{\end{tabular}}
\newcommand{\no}{\nonumber}
\newcommand{\non}{\nonumber\\}
\newcommand{\lbr}{\left [}
\newcommand{\rbr}{\right ]}
\newcommand{\lb}{\left \{ }
\newcommand{\rb}{\right \}}
\newcommand{\lp}{\left (}
\newcommand{\rp}{\right )}

\newcommand{\rd}{\right.}
\newcommand{\ve}[1]{{\bf #1}}
\newcommand{\vk}{\ve{k}}
\newcommand{\rhok}{{\rho_{\vk}}}
\newcommand{\rhomk}{{\rho_{-\vk}}}
\newcommand{\cB}{{\cal B}}
\newcommand{\slip}{\mathop{\sum\nolimits'}\limits}

\begin{document}

\binoppenalty=10000
\relpenalty=10000

\begin{center}
\textbf{\Large{EQUATION OF STATE OF A CELL FLUID MODEL WITH ALLOWANCE FOR GAUSSIAN FLUCTUATIONS OF THE ORDER PARAMETER }}
\end{center}

\vspace{0.3cm}

\begin{center}
I.V.~Pylyuk and O.A.~Dobush 
\end{center}

\begin{center}
Institute
for Condensed Matter Physics of the National Academy of Sciences of
Ukraine \\ 1, Svientsitskii Str., 79011 Lviv, Ukraine
\end{center}

 \vspace{0.5cm}

\small The paper is devoted to the development of a microscopic description of the critical behavior of a cell fluid model with allowance for the contributions from collective variables with nonzero values of the wave vector. The mathematical description is performed in the supercritical temperature range ($T>T_c$) in the case of a modified Morse potential with additional repulsive interaction. The method, developed here for constructing the equation of state of the system by using the Gaussian distribution of the order parameter fluctuations, is valid beyond the immediate vicinity of the critical point for a wide range of density and temperature. The pressure of the system as a function of chemical potential and density is plotted for various fixed values of the relative temperature, both with and without considering the above-mentioned contributions. Compared with the results of the zero-mode approximation, the insignificant role of these contributions is indicated for temperatures $T>T_c$. At $T<T_c$, they are more significant.

\vspace{0.5cm}

PACs: 05.70.Ce, 64.60.F-, 64.70.F-


\normalsize

\section{Introduction}

We dedicate this paper to the 75th anniversary of Academician L.A. Bulavin, an outstanding Ukrainian physicist, who made a remarkable contribution to the development of experimental base in the field of phase transitions and critical phenomena in simple and multicomponent liquid systems \cite{bul1,bul2,bul3,bul4}.

A large number of works are devoted to the description of the critical properties of liquid systems, a detailed bibliography of which can be found, for example, in the books {\cite{hm113,gspmo110,amo191,bmo102}} and review papers \cite{ss186,as100}.  For decades the interest in this problem persists because liquid near its critical points is the most convenient object, which simulates a class of systems with a large number of strongly interacting degrees of freedom \cite{gspmo110,amo191}. On the other hand, due to their specific properties, critical fluids are frequently used in various technological processes \cite{ztc106}. In this regard, building the equation of state of critical fluids becomes an important applied task.

The main difficulty of successive theoretical calculation of the equation of state by methods of statistical physics is the need for correctly taking into account the complex structure of interparticle interaction. Therefore, when calculating, it is necessary to use simplified models, the scope of which is limited and is either established in each case or based on internal characteristics of the model, or by comparison with more accurate solutions and experimental results.

In our previous studies \cite{kkd118,kd117,kdp117}, we proposed a cell fluid model, which we used to describe a first-order phase transition applying different types of interaction potentials. In this work, a modified Morse potential with a term describing a soft wall repulsion is used to calculate the equation of state of a cell fluid model.

We consider a system of $N$ interacting particles in the volume $V$ conditionally divided into $N_v$ cells ($ V = vN_v $, $ v = c^3$  is the volume, and $ c $ is the linear size of a cell) \cite{kdp117,kpd118}. Note that, in contrast to the lattice gas model (where it is assumed that the cell may or may not contain only one particle), in this approach, the cell may contain more than one particle. The interaction potential of such a cell fluid model
\be\label{1ma}
U(r) = C_H \lb Ae^{-n_0(r-R_0)/\alpha} + e^{-\gamma(r-R_0)/\alpha}
	- 2 e^{-(r-R_0)/\alpha}\rb,
\ee
along with the repulsive and attractive interactions (the second and third terms that form a Morse-type potential), includes an additional repulsive interaction (the first term) \cite{kd120}. Here $ R_0 $ corresponds to the minimum of the function $ U(r) $, $ r $ is the distance between particles, $ \alpha $ is the radius of effective interaction, $ \gamma $, $ n_0 $ are the parameters of the model, $ A = ( 2- \gamma) / n_0 $, $ C_H = Dn_0 / (n_0 + \gamma-2) $, $ D $ determines the dissociation energy. The quantities $ R_0 $, $ \alpha $, and $ D $ are specific to a particular physical system. In case of sodium (Na) we have \cite{lkg167,singh}
\be
R_0 = 5.3678 \mbox{ \normalfont\AA} , \quad
	1/\alpha = 0.5504 {\mbox{ \normalfont\AA}}^{-1}, \quad R_0/\alpha = 2.9544, \quad	D = 0.9241 \cdot 10^{-13} ergs.
\label{2ma}
\ee

The following expressions form a lattice analog of the potential (\ref{1ma})
\begin{align}\label{3ma}
& \tilde\Phi^{(r)}_{l_{12}} = C_H e^{-\gamma(l_{12}-c)/(\alpha_R c)}, \non
& \tilde\Phi^{(a)}_{l_{12}} = 2 C_H e^{-(l_{12}-c)/(\alpha_R c)}, \non
& \tilde \Psi_{l_{12}} = C_H A e^{-n_0(l_{12}-c)/(\alpha_R c)},
\end{align}
where $l_{12}=|\ve{l}_{1}-\ve{l}_{2}|$ is the distance between cells $\ve{l}_{1}$ and
$\ve{l}_{2}$ , $\alpha_R=\alpha/R_0$ is the dimensionless quantity. In what follows, we will use Fourier transforms of interaction potentials (\ref{3ma}), which have the form

\begin{align}\label{4ma} 
&\Phi^{(r)}(k) = C_H 8\pi e^{\gamma/\alpha}
	\left( \frac{\alpha}{\gamma}\right)^{\!\! 3}
	\lbr 1 + \left( \frac{\alpha}{\gamma}\right)^{\!\! 2} c^2 k^2 \rbr^{\!\! -2}, \non
& \Phi^{(a)}(k) = C_H 16\pi e^{1/\alpha} \alpha^3
	\lbr 1 + \alpha^2 c^2 k^2 \rbr^{-2}, \\
& \Psi(k) = C_H A 8\pi e^{n_0/\alpha}
	\left( \frac{\alpha}{n_0}\right)^{\!\! 3}
	\lbr 1 + \left( \frac{\alpha}{n_0}\right)^{\!\! 2} c^2 k^2 \rbr^{\!\! -2}. \no
\end{align}
Hereinafter, to simplify the notation, $ \alpha $ should be understood as the quantity $ \alpha_R $, and $ c $ is $ c_R = c / R_0 $. Easy to see that
\begin{align}\label{5ma}
& \Phi^{(a)}(0) = B \Phi^{(r)}(0), \quad B = 2 \gamma^3 e^{(1-\gamma)/\alpha}, \non
& \Psi(0) = A_\gamma \Phi^{(r)}(0), \quad A_\gamma = A e^{(n_0-\gamma)/\alpha} \left( \gamma / n_0\right)^3.
\end{align}

This work logically complements the previous studies \cite{kd120}, devoted to the theoretical description of the first-order phase transition in the cell fluid model with potential (\ref{1ma}). In \cite{kd120}, the equation of state of the system in terms of chemical potential-temperature and density-temperature is calculated in the zero-mode approximation. Contributions from the collective variables $ \rhok $ with nonzero values of the wave vector were not taken into account when obtaining the equation of state of the cell fluid model. The purpose of this work is to develop a technique for constructing the equation of state of the model, taking into account these contributions, compare the results obtained for system pressure taking into account, and leaving out these contributions, as well as assess the influence of the above-mentioned contributions on calculations.
The equation of state obtained in this paper is not applicable in the immediate vicinity of the critical point ($|\tau|<\tau^*\sim 10^{-2}$). In \cite{kpd118,p120}, one finds a description of the method that gives the equation of state in this narrow neighborhood of $ T_c $ taking into account non-Gaussian fluctuations. A significant role in the development of this technique played works \cite{kpp506,kpu196,KR_2012,K_2009,kp394}, devoted to describing the behavior of a three-dimensional ising-like system near $ T_c $ using the quartic measure density (the $\rho^4$ model).

\section{The grand partition function of the system}

The grand partition function of the cell fluid model in the approximation of the $\rho^4$ model is given by the formula \cite{kpd118}
\begin{align} \label{6ma}
& \Xi = 2^{(N_v-1)/2} g_v  e^{N_vE_\mu} \int (d\rho)^{N_v}
	\exp \Biggl[ M N_v^{1/2} \rho_0 + \frac{1}{2} \sum_{\vk\in \cB_c} \tilde D(k) \rhok\rhomk + \non
& + \frac{g_4}{24} \frac{1}{N_v}
	\sum_{\substack{\vk_1,\ldots,\vk_4 \\ \vk_{i}\in \cB_c}}
	\rho_{\vk_1}\cdots\rho_{\vk_4} \delta_{\vk_1+\cdots+\vk_4}\Biggr]
\end{align}
with the following notations:
\begin{align} \label{7ma}
& E_\mu = g_0 \! - \! \frac{\beta\tilde\mu^2}{2W(0)} \! + \!
	n_c \! \left( \! g_1 \! + \! \frac{\tilde\mu}{W(0)} \! \right) \! + \!
	\frac{n_c^2}{2} \tilde D(0) \! + \! \frac{g_3^4}{8g_4^3}, \non
& M = \tilde\mu / W(0) + g_1 + n_c \tilde D(0) -
	\frac{1}{6} \frac{g_3^3}{g_4^2}, \non
& \tilde D(k) = \tilde g_2 - 1 /(\beta W(k)), \non
& \tilde g_2 = g_2 - \frac{1}{2} \frac{g_3^2}{g_4}, \quad
	n_c = - g_3 / g_4.
\end{align}
The quantity $\tilde\mu$ characterizes the chemical potential, and $\beta \!=\!1/(kT)$ is the inverse temperature. Using the expressions (\ref{4ma}) and (\ref{5ma}), we can write the total Fourier transform of the effective potential $ W(k) = \Phi^{(a)}(k) - \Phi^{(r)}(k) - \Psi(k) +
\frac{\beta_c}{\beta} \chi_0 \Phi^{(r)}(0) + \frac{\beta_c}{\beta} \Psi(0)$ in the form
\begin{align}\label{8ma}
& W(k) = \Phi^{(r)}(0) \Biggl\{ \frac{B}{\bigl[ 1 + \alpha^2 c^2 k^2 \bigr]^2} -
	\frac{1}{\bigl[ 1 + \bigl( \frac{\alpha}{\gamma}\bigr)^2 c^2 k^2 \bigr]^2} - \frac{A_\gamma}{\bigl[ 1 + \bigl( \frac{\alpha}{n_0}\bigr)^2 c^2 k^2 \bigr]^2} +
	\bigl( 1 + \tau \bigr) \bigl( \chi_0 + A_\gamma \bigr) \Biggr\},
\end{align}
where $\chi_0$ is some constant quantity, $\tau = (T - T_c) / T_c$ is the relative temperature, $T_c$ is the critical temperature.
At $k = 0$, we obtain
\be\label{9ma}
W(0) = \Phi^{(r)}(0) \left[ B - 1 + \chi_0 + \tau (\chi_0 + A_\gamma) \right].
\ee
The coefficients $g_n$, which appear in (\ref{7ma}), are given by the formulas  \cite{kdp117,kd116}
\begin{align}\label{10ma}
& g_0 = \ln T_0, \quad g_1 = T_1/T_0, \quad g_2 = T_2/T_0 - g_1^2, \non
& g_3 = T_3/T_0 - g_1^3 - 3 g_1 g_2, \non
& g_4 = T_4/T_0 - g_1^4 - 6 g_1^2 g_2 - 4 g_1 g_3 - 3 g_2^2.
\end{align}
The special functions $T_n(p,\alpha^*)$ are as follows:
\be\label{11ma}
T_n(p,\alpha^*) = \sum_{m=0}^{\infty} \frac{(\alpha^*)^m}{m!} m^n e^{-pm^2}.
\ee
Here $p = \beta_c \Phi^{(r)}(0) [\chi_0 + A_\gamma]/2$, and $\alpha^*\sim v$
(see \cite{kd120}). The expression of the grand partition function (\ref{6ma}) also includes the quantity
\be\label{12ma}
g_v = \prod_{\vk\in \cB_c} (2\pi \beta W(k))^{-1/2},
\ee
which is a function of the inverse temperature $\beta$ and the Fourier transform of the effective potential $W(k)$. The wave vector $\vk$ runs all the values inside the Brillouin zone
\begin{align*}
& \cB_c = \Big\{ \vk \!\!=\!\! (k_1, k_2, k_3)\Big| k_{i} \!\!=\!\!
	- \frac{\pi}{c} + \frac{2\pi}{c}\frac{n_i}{N_a}; \, n_i=1,2,\ldots,N_a; i=1,2,3; N_v=N_a^3 \Big\}.
\end{align*}

Singling out terms with $k=0$ from the sums over $\vk$ in (\ref{6ma}), we obtain
\be
\Xi = 2^{(N_v-1)/2} \Xi' g_v \exp [N_v (E_\mu + E(\bar\rho_0))],
\label{13ma}
\ee
where
\be
E(\bar\rho_0) = \bar M \bar\rho_0 + \frac{1}{2} \tilde D(0) \bar\rho_0^2 -
\frac{a_4}{24} \bar\rho_0^4,
\label{14ma}
\ee
moreover the coefficient $a_4 = -g_4 > 0$. The quantity $\bar\rho_0$ can be determined from the condition
$\partial E(\rho_0) / \partial \rho_0 = 0 \big|_{\rho_0=\bar\rho_0}$,
which leads to the equation
\be
\bar M + \tilde D(0) \bar\rho_0 - \frac{a_4}{6} \bar\rho_0^3 = 0.
\label{15ma}
\ee
Here $\bar M$ is the chemical potential, which corresponds to an extremum of the function $E(\bar\rho_0)$.
At $T>T_c$, the real solution of (\ref{15ma}) takes the form
\begin{align}
& \bar\rho_0 = \lp \frac{3\bar M}{a_4} + \sqrt{Q_t} \rp^{1/3} -
\lp - \frac{3\bar M}{a_4} + \sqrt{Q_t} \rp^{1/3}, \non
& Q_t = \lp - \frac{2\tilde D(0)}{a_4} \rp^3 + \lp \frac{3\bar M}{a_4} \rp^2. \no
\end{align}
For the component $\Xi'$ of the grand partition function, we have
\begin{align}\label{16ma}
& \Xi' = \int (d\rho)^{N_v-1}
	\exp \Biggl[ - \frac{1}{2} \slip_{\vk\in \cB_c} \tilde d(k) \rhok\rhomk - \frac{a_4}{4} \frac{1}{N_v} \rho_0^2
	\slip_{\vk\in \cB_c} \rhok\rhomk - \non
& - \frac{a_4}{24} \frac{1}{N_v}
	\slip_{\substack{\vk_1,\ldots,\vk_4 \\ \vk_{i}\in \cB_c}}
	\rho_{\vk_1}\cdots\rho_{\vk_4} \delta_{\vk_1+\cdots+\vk_4} \Biggr],
\end{align}
where $\tilde d(k) = - \tilde D(k)$. The prime next to the sum sign means that the term with $ k = 0 $ is missing.

Carrying out integration in (\ref{16ma}) with respect to the variables $ \rhok $ with $ k \neq 0 $ by using the Gaussian distribution of fluctuations as the basis one, we obtain the following expression in a zero-order approximation:
\be
\Xi' = \prod_{k\neq 0}^{\cB_c} (\pi / \tilde d_A(k))^{1/2}.
\label{17ma}
\ee
Here
\be
\tilde d_A(k) = \tilde d(k) + A',
\label{18ma}
\ee
and
\be
\tilde d(k) = 1 /(\beta W(k)) - \tilde g_2.
\label{19ma}
\ee
The quantity $A'$ is given by the formula
\be
A' = \frac{a_4}{4} (1 + 2<\rho_0^2>) N_v^{-1}
\slip_{\vk\in \cB_c} <\rhok\rhomk>.
\label{20ma}
\ee
It should be noted that in the calculation process, we took into account the approximation
\begin{align}\label{21ma}
& \slip_{\substack{\vk_1,\ldots,\vk_4 \\ \vk_{i}\in \cB_c}}
	\rho_{\vk_1}\cdots\rho_{\vk_4} \delta_{\vk_1+\cdots+\vk_4} \approx
	3 \slip_{\vk_1\in \cB_c} <\rho_{\vk_1}\rho_{-\vk_1}> \slip_{\vk_2\in \cB_c} \rho_{\vk_2}\rho_{-\vk_2}.
\end{align}
Both the expression (\ref{20ma}) and the relations
\begin{align}\label{22ma}
& <\rhok\rhomk> = \frac{1}{\tilde d_A(k)}, \non
& <\rho_0^2> = \left\{
	\begin{array}{ll}
		0, &  T\geq T_c, \\
		- \frac{6 \tilde d(0)}{a_4}, & T<T_c
	\end{array} \rd
\end{align}
allows us to find the equation for $ A '$. It is obtained by a transition to the spherical Brillouin zone and replacing the summation over the wave vectors by integration with respect to $ \vk \in \cB_c $. The equation for $ A'$ has the form
\be
A' = A'_\tau \int \limits_{0}^{\cB_c} \frac{\beta W(k) k^2}
{1 - \beta W(k) (\tilde g_2 - A')} dk,
\label{23ma}
\ee
where
\be
A'_\tau = \left\{
\begin{array}{ll}
	\frac{3}{4} a_4 \frac{c^3}{\pi^3}, &  T\geq T_c, \\
	\frac{3}{4} a_4 \frac{c^3}{\pi^3} \lp 1 - \frac{12 \tilde d(0)}{a_4} \rp, & T<T_c.
\end{array} \rd
\label{24ma}
\ee
The quantity $A'$ is included in the coefficient $\tilde d_A(k)$ (\ref{18ma}), through which the component of the grand partition function $\Xi'$ (\ref{17ma}) is expressed.
The component $\Xi'$ and the quantity $g_v$ (\ref{12ma}) are contained in the expression for $\Xi$ (\ref{13ma}). Let us write the expressions for the logarithms $ \Xi'$ and $ g_v $, which will be needed to calculate the equation of state of the system.

Taking into account the relation
\be
\frac{1}{2} \slip_{\vk\in \cB_c} \! \ln\tilde d_A(k) \! = \!
\frac{3}{2} N_v \frac{c^3}{\pi^3} \! \int \limits_{0}^{\cB_c} \!\!
k^2 \ln \! \lbr \frac{1}{\beta W(k)} \! - \! \tilde g_2 \! + \! A' \rbr \! dk,
\label{25ma}
\ee
we find the following expression for $\ln\Xi'$:
\be
\ln\Xi' = N_v {L_{\Xi'}(T)}.
\label{26ma}
\ee
Here
\be
{L_{\Xi'}(T)} \!\! = \!\! \frac{1}{2} \ln\pi \! - \!
\frac{3}{2} \frac{c^3}{\pi^3} \!\! \int \limits_{0}^{\cB_c} \!\!
k^2 \ln \!\! \lbr \frac{1}{\beta W(k)} \! - \! \tilde g_2 \! + \! A' \rbr \!\! dk.
\label{27ma}
\ee
For $\ln g_v$, we finally get
\be
\ln g_v = N_v {L_{g_v}(T)},
\label{28ma}
\ee
where
\be
{L_{g_v}(T)} \! = \! - \frac{1}{2} \ln (2 \pi) \! - \!
\frac{3}{2} \frac{c^3}{\pi^3} \! \int \limits_{0}^{\cB_c} \!\!
k^2 \ln \! \lbr \frac{W(k)}{kT} \rbr \! dk.
\label{29ma}
\ee

\begin{figure}[h!]
	\vskip1mm
	\includegraphics[width=0.45\textwidth]{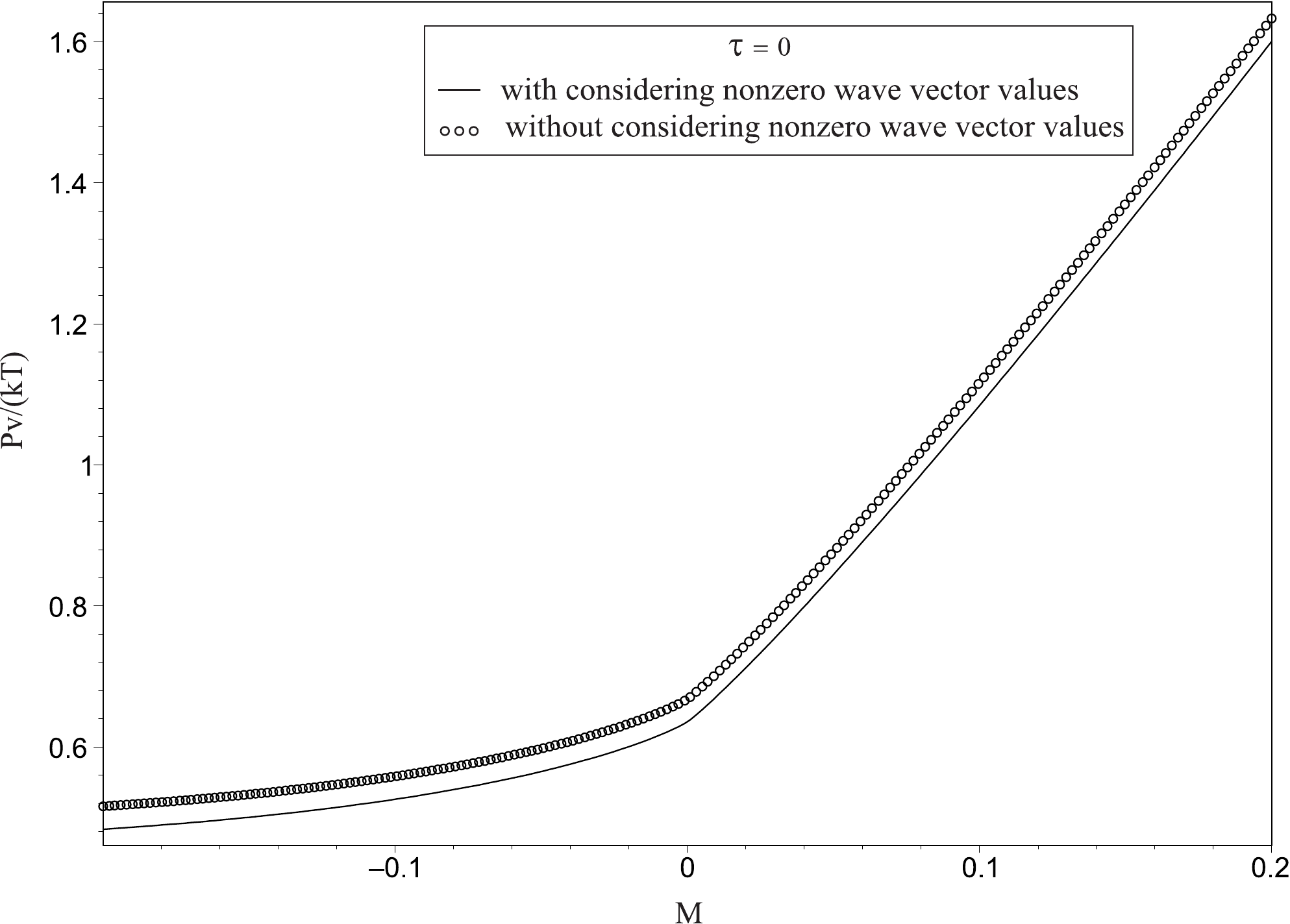}
	\hfill
	\includegraphics[width=0.45\textwidth]{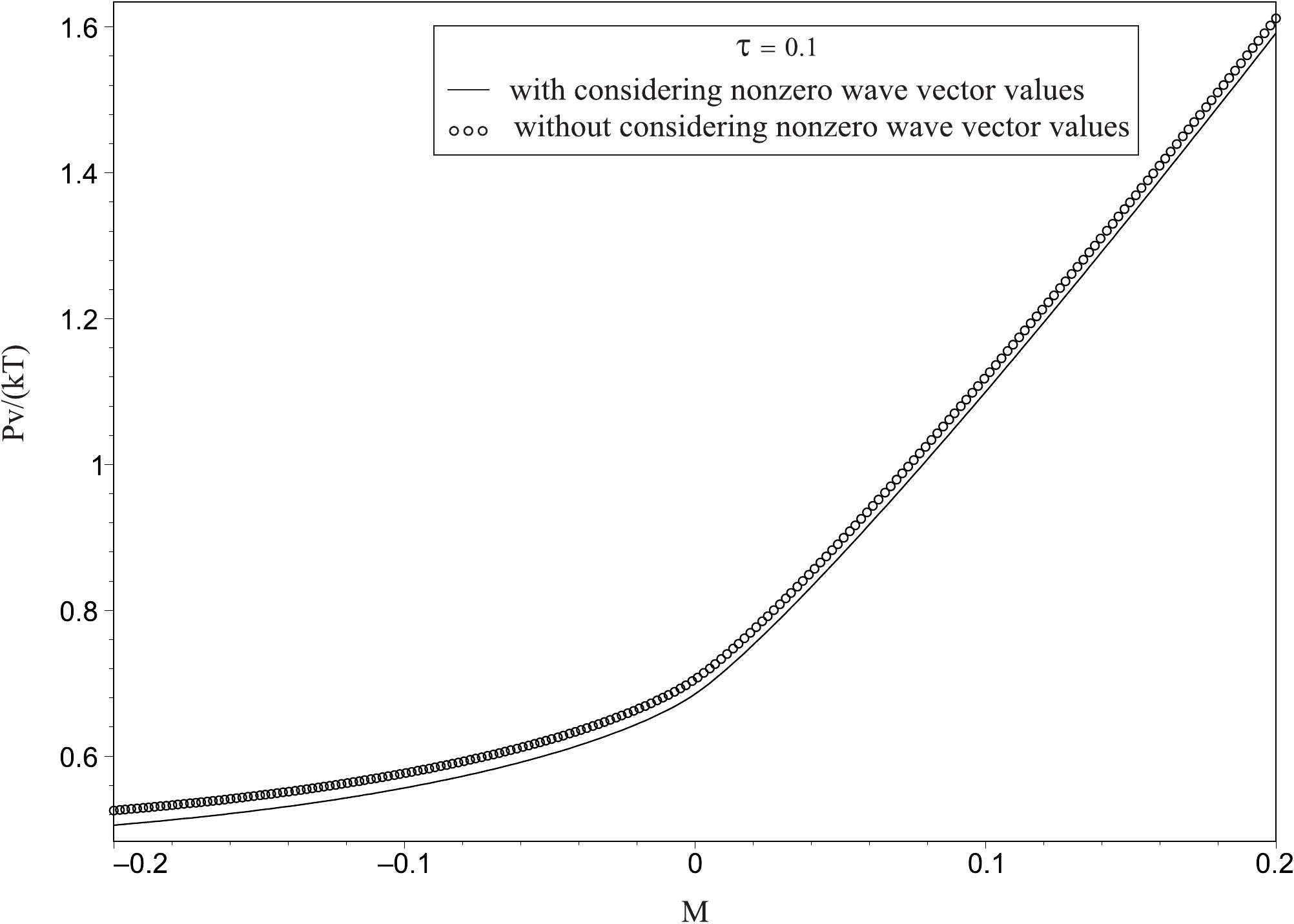}
	\\ [0.3cm]
	\includegraphics[width=0.45\textwidth]{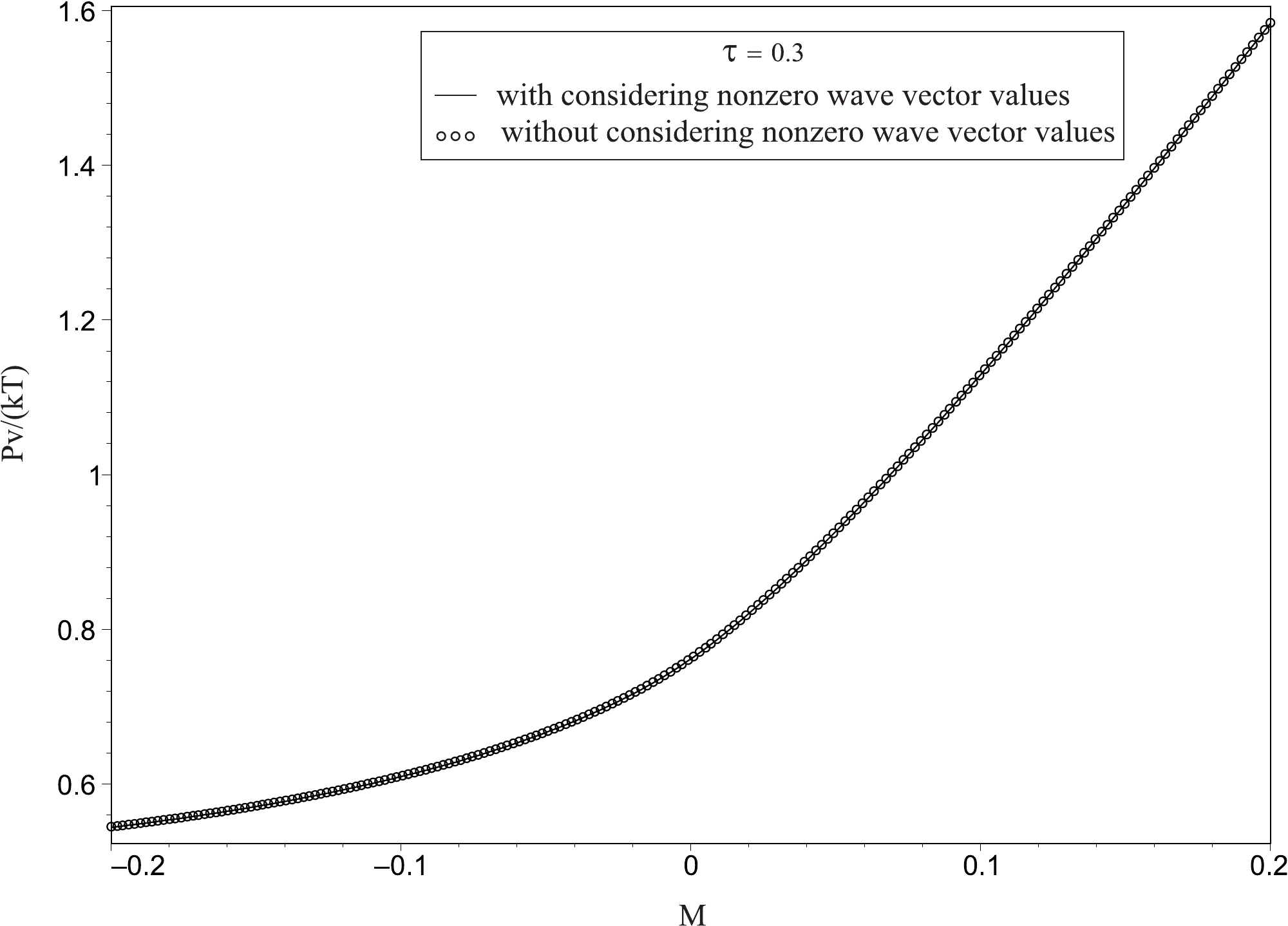}
	\hfill
	\includegraphics[width=0.45\textwidth]{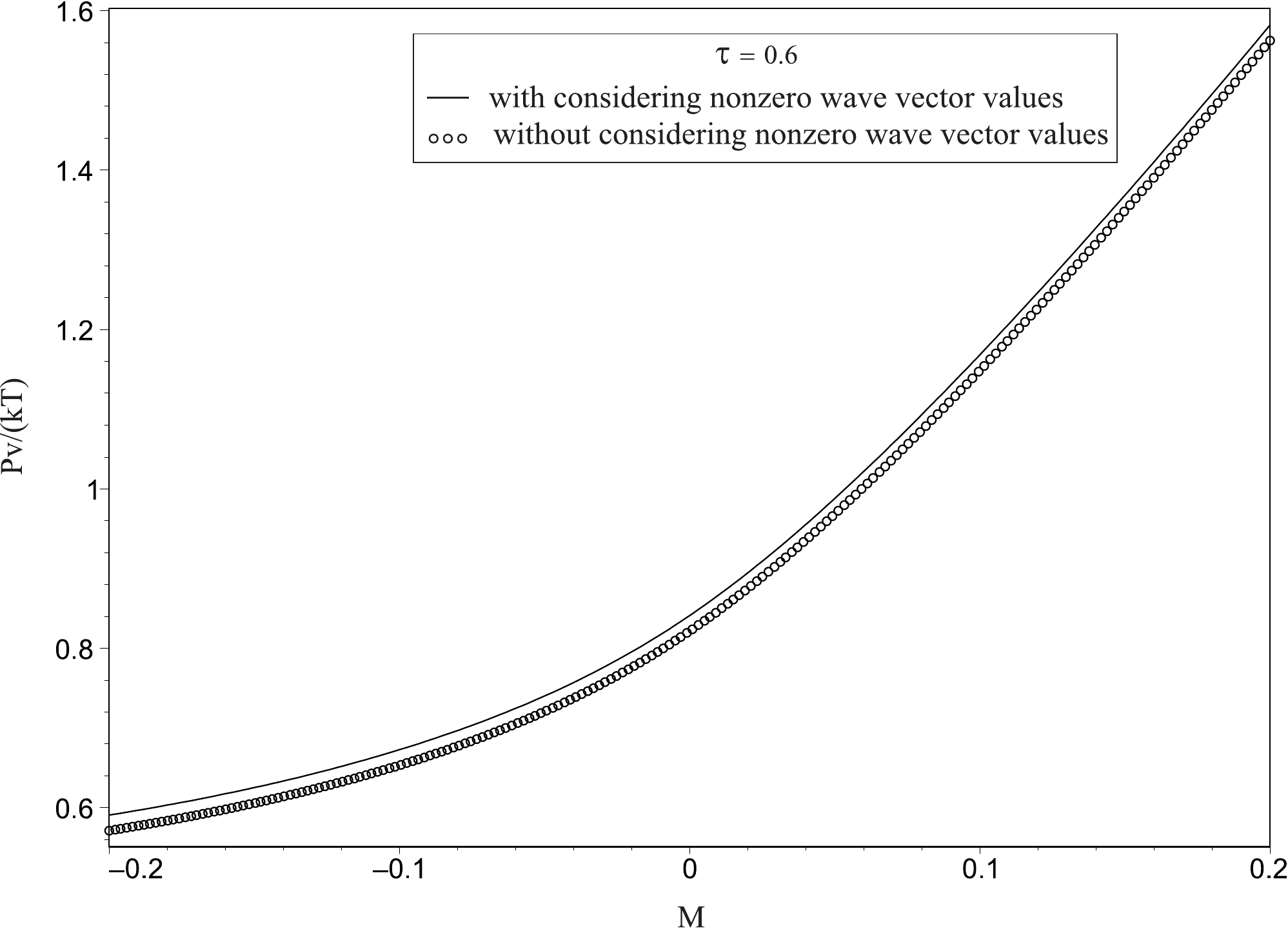}
	\vskip-3mm\caption{Dependence of $Pv/(kT)$ on the chemical potential $M$ for various values of the relative temperature $\tau$
		without taking into account the contributions from the quantities ${L_{\Xi'}(T)}$ and ${L_{g_v}(T)}$ (circles)
		and taking into account the mentioned contributions (solid curves)}
	\label{fig_1ma}
\end{figure}

The expressions (\ref{26ma}) -- (\ref{29ma}) allow you to find the sum
$\ln\Xi' + \ln g_v$. This sum together with the term $N_v \ln 2/2$ appearing in $\Xi$ (\ref{13ma}) from the first multiplier will additionally be included in the equation of state of the cell fluid model, which was obtained in \cite{kd120} in the zero-mode approximation (the mean-field approximation) for the potential (\ref{1ma}). It characterizes the contribution from collective variables with nonzero values of the wave vector. The values of the quantities
$\tilde d(0)$, $A'$, as well as ${L_{\Xi'}}$ (\ref{27ma}),
${L_{g_v}}$ (\ref{29ma}) and their sum ${L_{\Xi'g_v}} = {L_{\Xi'}} + {L_{g_v}}$
are given in the Table~\ref{tab_1ma} for various $\tau$.
	\begin{table}[b]
	\noindent\caption{Values of $\tilde d(0)$, $A'$, and contributions to the pressure of the system from the quantities $\Xi'$ and $g_v$ at different temperatures~$\tau$}\vskip3mm\tabcolsep4.5pt
	\label{tab_1ma}
	\noindent{\footnotesize
			\begin{tabular}{|c|c|c|c|c|c|c|}
			\hline%
			\multicolumn{1}{|c}{\rule{0pt}{5mm}$\tau$}%
			& \multicolumn{1}{|c}{$\tilde d(0)$}
			& \multicolumn{1}{|c}{$A'$}
			& \multicolumn{1}{|c}{${L_{\Xi'}}$}
			& \multicolumn{1}{|c}{${L_{g_v}}$}
			& \multicolumn{1}{|c}{${L_{\Xi'g_v}}$}
			& \multicolumn{1}{|c|}{$\!\!{L_{\Xi'g_v}}\!\!\!+\!\!\frac{\ln 2}{2}\!\!$}\\[2mm]%
			\hline%
			\rule{0pt}{5mm}$-0.5$&$-0.081$&0.247&0.459&$-0.759$&$-0.300$&0.047\\%
			$-0.3$&$-0.040$&0.194&0.733&$-0.977$&$-0.244$&0.103\\%
			$-0.1$&$-0.011$&0.097&0.929&$-1.061$&$-0.132$&0.215\\%
			0&0&0.030&1.054&$-1.087$&$-0.033$&0.314\\%
			0.1&0.010&0.031&1.086&$-1.107$&$-0.021$&0.326\\%
			0.3&0.026&0.033&1.135&$-1.135$&0.000&0.346\\%
			0.5&0.039&0.035&1.169&$-1.155$&0.014&0.360\\[2mm]%
			\hline
		\end{tabular}
	}
\end{table}

Hereinafter, the calculations are performed for the following set of parameters:
\begin{align}\label{30ma}
& p = 1.0, \quad \alpha^* = 5.0, \quad v = 1.0; \non
& \gamma = 1.330, \quad \chi_0 = 0.070.
\end{align}
The value of the ratio $R_0/\alpha$ is given in (\ref{2ma}). For the quantities $n_0$ and $A_\gamma$, taking into account (\ref{30ma}), we will have
\be
n_0 = 1.541, \quad A_\gamma = 0.521.
\label{31ma}
\ee

Let us go over to calculating the equation of state of the system.

\section{Equation of state of the model with allowance for
	Gaussian fluctuations}

\begin{figure}[h!]
	\vskip1mm
	\includegraphics[width=0.45\textwidth]{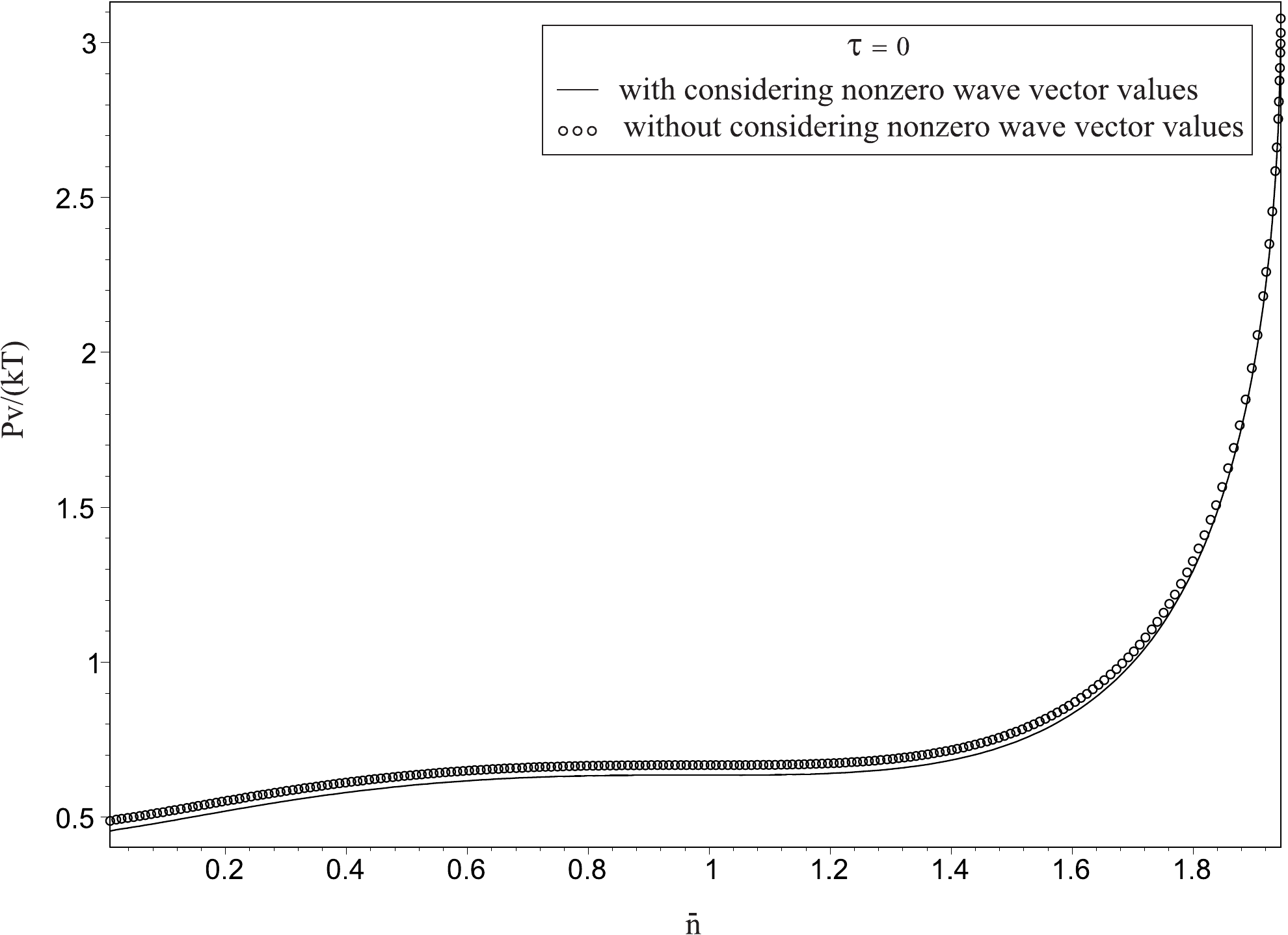}
	\hfill
	\includegraphics[width=0.45\textwidth]{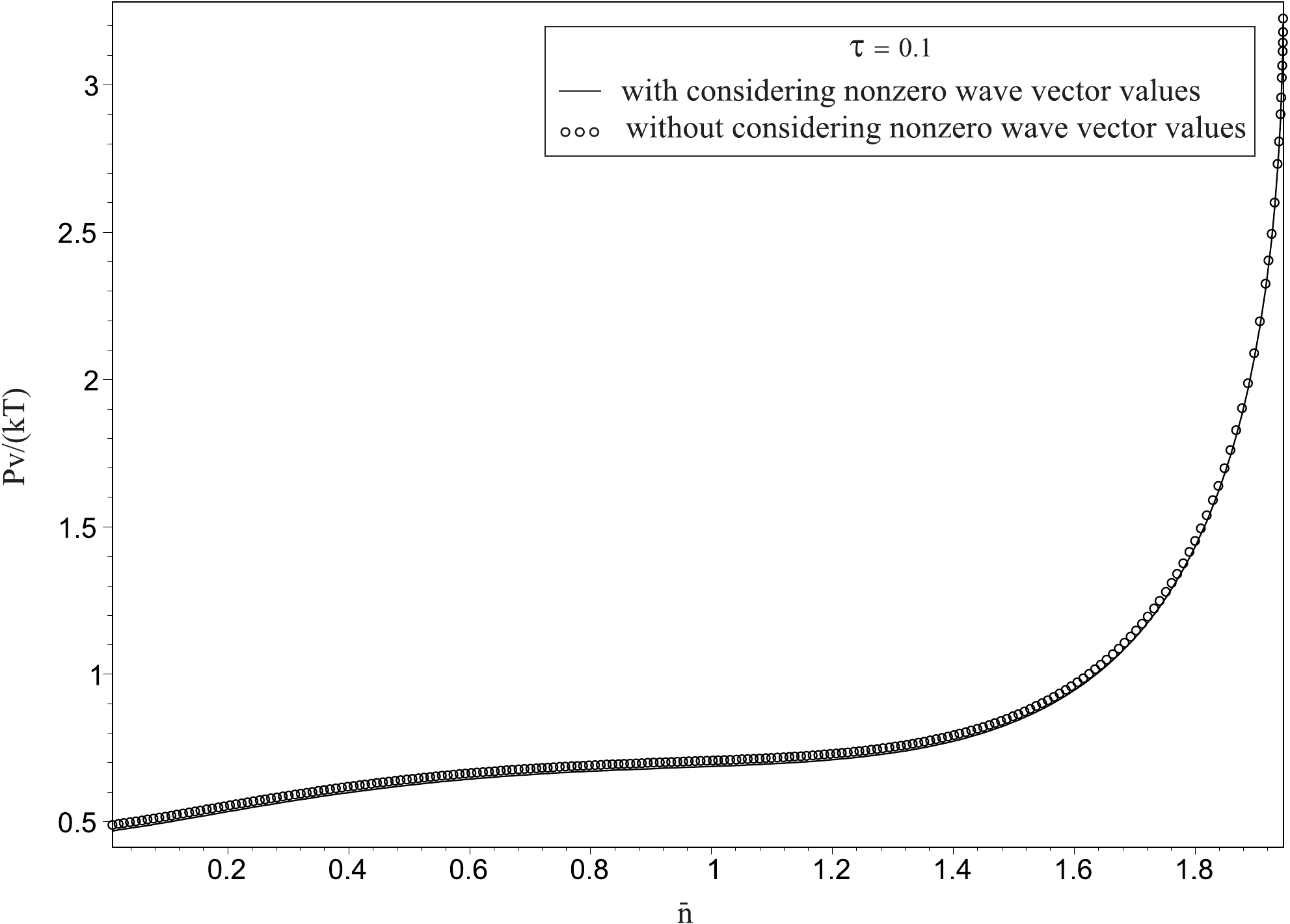}
	\\ [0.3cm]
	\includegraphics[width=0.45\textwidth]{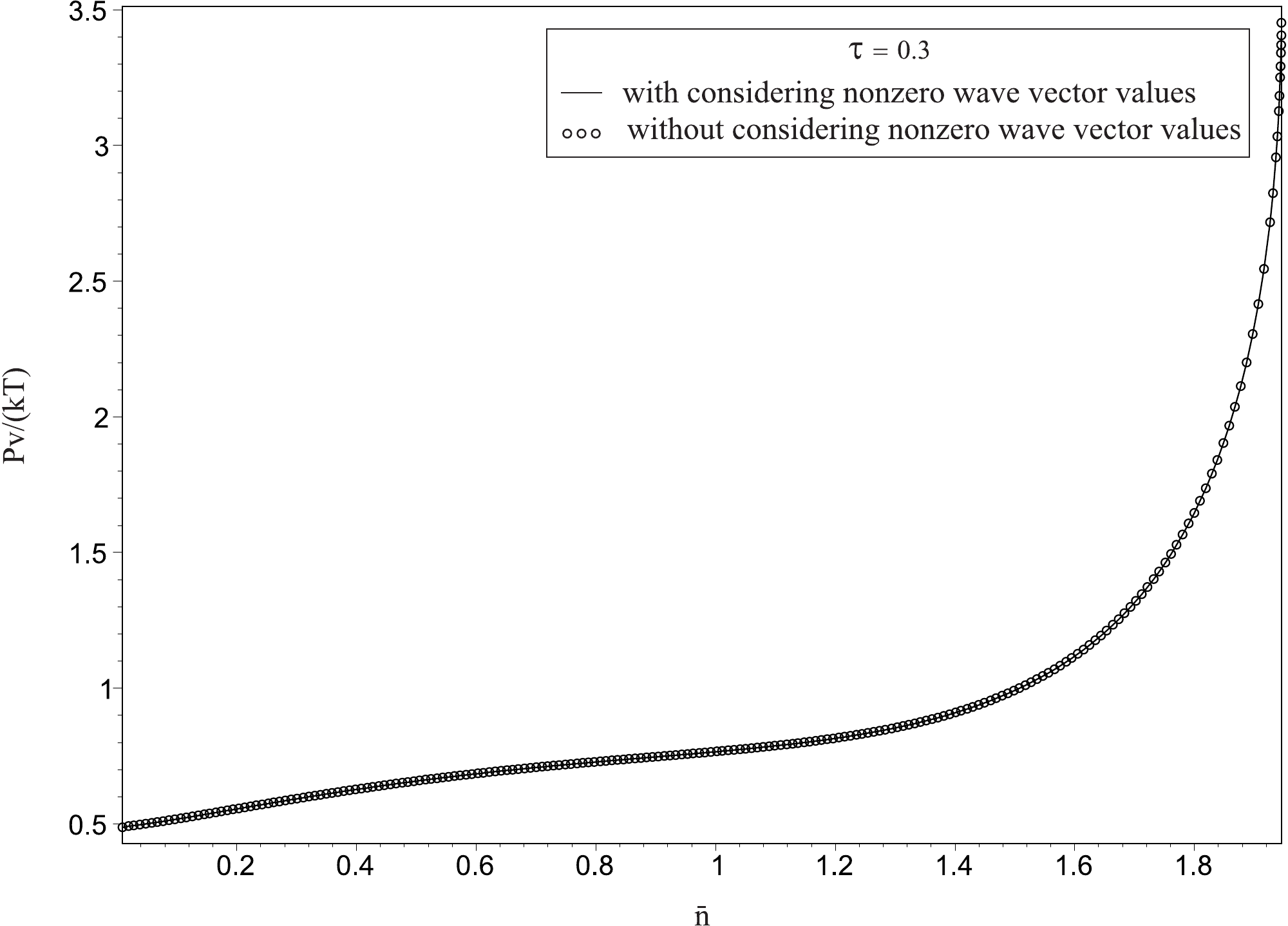}
	\hfill
	\includegraphics[width=0.45\textwidth]{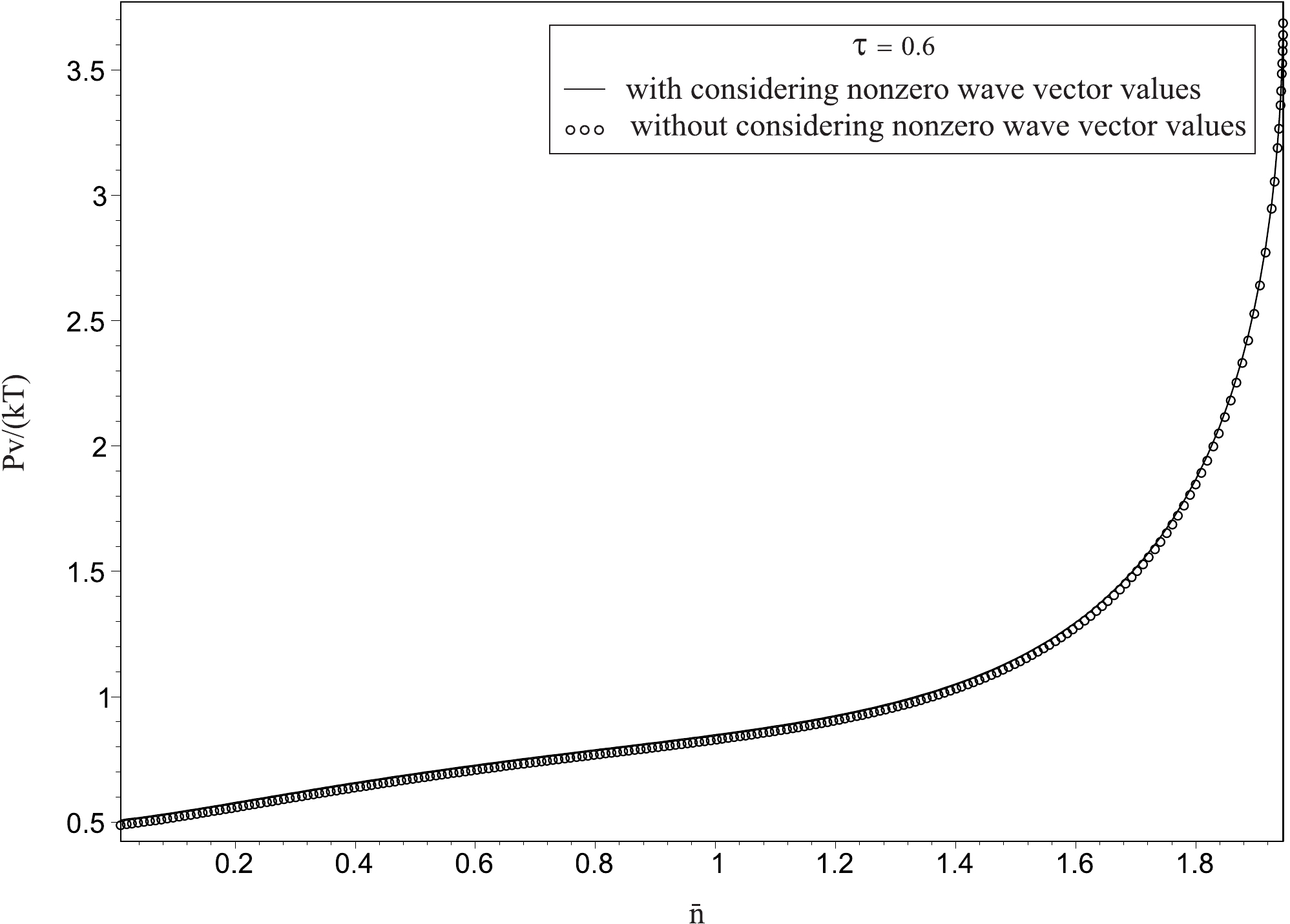}
	\vskip-3mm\caption{Pressure as a function of the average density for various values of the relative temperature.
		The results were obtained both taking into account the contributions from collective variables with nonzero values of the wave
		vector (solid curves) and without them (circles)}
	\label{fig_2ma}
\end{figure}

Applying the well-known relation
\be
PV = k T \ln\Xi
\label{32ma}
\ee
and the formula (\ref{13ma}), we find an explicit form of the equation of state in the case of $T>T_c$ that is either
\begin{align}\label{33ma}
& \frac{P V}{k T} = \ln\Xi' + \ln g_v +
	N_v \biggl[ \frac{1}{2} \ln 2 + E_\mu(M,T) +  \bar M \bar\rho_0 + \frac{1}{2} \tilde D(0) \bar\rho_0^2 -
	\frac{a_4}{24} \bar\rho_0^4 \biggr]
\end{align}
or
\begin{align}\label{34ma}
& \frac{P v}{k T} = {L_{\Xi'}(T)} + {L_{g_v}(T)} +
	\frac{1}{2} \ln 2 + E_\mu(M,T) + \bar M \bar\rho_0 + \frac{1}{2} \tilde D(0) \bar\rho_0^2 -
	\frac{a_4}{24} \bar\rho_0^4.
\end{align}
Here
\begin{align}
& E_\mu (M,T)  =  g_0  - \frac{\beta W(0)}{2} \lp \frac{\tilde\mu}{W(0)} \rp^2 +
\bar M n_c  -\frac{n_c^2}{2}  \tilde D(0) - \frac{1}{24} \frac{g_3^4}{g_4^3}, \non
& \frac{\tilde\mu}{W(0)} = \bar M - g_1 - n_c \tilde D(0) +
\frac{1}{6} \frac{g_3^3}{g_4^2}.
\label{35ma}
\end{align}
The expression for the pressure (\ref{34ma}) at $ T> T_c $ is a monotonically increasing function of $ \bar M $ [the variable $ M $ in the formalism of the grand canonical ensemble in coordinates ($ \tau, M $)] (see Fig.~\ref{fig_1ma}).

The equation (\ref{34ma}) describes the dependence of the pressure on temperature and chemical potential. Let us now go over to obtaining the dependence of the pressure on temperature and density.

Taking into account the expression of the grand partition function (\ref{13ma}), we can find the average number of particles
\be
\bar N = \frac{\partial \ln\Xi}{\partial \beta\mu}
\label{36ma}
\ee
or the average density
\be
\bar n = \frac{\bar N}{N_v}.
\label{37ma}
\ee
Considering the relations (\ref{14ma}) and (\ref{35ma}), we get the equation
\be
\bar n = n_g - \bar M + \tilde g_2 \gamma_\tau \bar\rho_0
\label{38ma}
\ee
connecting the density of particles $\bar n$ with the chemical potential $\bar M$.
Here
\begin{align}\label{39ma}
& n_g = g_1 + n_c \tilde g_2 - \frac{1}{6} g_3^3 / g_4^2, \non
& \gamma_\tau = \frac{1 + \tau}{1 + \omega_0\tau}, \quad
	\omega_0 = \frac{\chi_0 + A_\gamma}{B - 1 + \chi_0}.
\end{align}
Substituting $\bar M$ from (\ref{15ma}) in the expression (\ref{38ma}), we arrive at the cubic equation for $\bar\rho_0$ (see \cite{kd120}). Among the three solutions of this equation, only one solution
\be
\rho_{02n} = - 2 \lp \frac{2\tilde g_2}{a_4} \rp^{1/2}
\cos \lp \alpha_n / 3 + \pi / 3 \rp
\label{40ma}
\ee
has a physical sense.
Here
\be
\alpha_n = \arccos \lp \frac{n_g - \bar n}{n_\varphi} \rp, \quad
n_\varphi = \frac{2}{3} \lp \frac{2\tilde g_2^3}{a_4}\rp^{1/2}.
\label{41ma}
\ee
The relations (\ref{38ma}) and (\ref{40ma}) allow us to express the chemical
potential $\bar M$ in terms of the average density $\bar n$. We will have
\be
\bar M = \rho_{02n} \tilde g_2 \gamma_\tau - (\bar n - n_g).
\label{42ma}
\ee

The equation of state of the cell fluid model at $ T> T_c $ in terms of temperature and density takes the form
\begin{align}\label{43ma}
& \frac{P v}{k T} = {L_{\Xi'}(T)} + {L_{g_v}(T)} +
	\frac{1}{2} \ln 2 + E_\mu(n,T) + \bar M \bar\rho_0 + \frac{1}{2} \tilde D(0) \bar\rho_0^2 -
	\frac{a_4}{24} \bar\rho_0^4.
\end{align}
The expression for $\bar M$ from (\ref{42ma}) should be substituted in (\ref{43ma}), as well as in  the solution $\bar\rho_0$ of the equation (\ref{15ma}) and in the relations
\begin{align}
& E_\mu(n,T) = g_0 - \frac{1}{2\tilde g_2 \gamma_\tau} \lp \frac{\tilde\mu}{W(0)}\rp^2 +
\bar M n_c - \frac{n_c^2}{2}  \tilde D(0) - \frac{1}{24} \frac{g_3^4}{g_4^3}, \non
& \frac{\tilde\mu}{W(0)} = \bar M - n_g + n_c \tilde g_2 \gamma_\tau.
\label{44ma}
\end{align}
The behavior of the pressure (\ref{43ma}) with increasing density $\bar n$ is shown in Fig.~\ref{fig_2ma} for various $\tau$.

\section{Conclusions}

In the temperature range $T> T_c$, the procedure for constructing the equation of state of the cell fluid model is developed
taking into account Gaussian fluctuations of the order parameter. The Gaussian fluctuation distribution is used as the basis one when calculating contributions from collective variables with nonzero values of the wave vector.

The contributions to the pressure of the system from the collective variables $\rhok$ with $k\neq 0$ are calculated for temperatures below and above the critical value of $T_c$
(see the quantities ${L_{\Xi'}}$, ${L_{g_v}}$, and ${L_{\Xi'g_v}}$ in Table~\ref{tab_1ma}). As is seen from Table~\ref{tab_1ma}, the quantity ${L_{\Xi'}}$ (\ref{27ma}) increases with the rise of temperature, and
${L_{g_v}}$ (\ref{29ma}) decreases. The total contribution of
${L_{\Xi'g_v}} = {L_{\Xi'}} + {L_{g_v}}$ to the pressure at temperatures $T<T_c$ ($\tau<0$) is more significant compared with the case of $T>T_c$ ($\tau>0$). This is evidenced by the magnitude (module) of the total contribution
${L_{\Xi'g_v}}$ for various $\tau$ (see Table~\ref{tab_1ma}).

The equation of state of the cell fluid model is obtained in terms of chemical potential-temperature and density-temperature with allowance for the abo\-ve-mentioned contributions.
The comparison of the behavior of system pressure in the presence and absence of these contributions, which is shown in Fig.~\ref{fig_1ma} and Fig.~\ref{fig_2ma}, indicates the insignificant role of contributions in the case of $T>T_c$.
For example, the magnitude of the total contribution to the system pressure from collective variables with nonzero values
of the wave vector (from $L_{\Xi'g_v}$) with changing density $\bar n$ at $\tau = 0.1$ (Fig.~\ref{fig_2ma}) does not exceed 4.5\%.
At $\tau = 0.3$ and $\tau = 0.6$ this magnitude of the contribution becomes even smaller.

Thus, it is established that in the region of supercritical temperatures ($T>T_c$), the inclusion of Gaussian fluctuations has a negligible effect on the equation of state of the cell fluid model.
Therefore, at $T>T_c$, the zero-mode approximation is enough for calculating the equation of state.

\vskip3mm \textit{The authors are sincerely grateful to Professor M.P. Kozlovskii for helpful advice and a detailed discussion of the results obtained.}

\end{document}